\documentclass[11pt]{article}
\usepackage{moriond,epsfig}

\def\be{\begin{equation}}
\def\ee{\end{equation}}
\def\bea{\begin{eqnarray}}
\def\eea{\end{eqnarray}}

\begin{document}
\vspace*{4cm}
\title{CONSTRAINTS ON CLUSTER PROPERTIES USING X-RAY \& OPTICAL DATA}

\author{ J.M DIEGO$^{1,2}$, E. MARTINEZ-CONZALEZ$^1$, J. L.
    SANZ$^1$, L. CAYON$^1$, J. SILK$^3$}

\address{   
   $^1$Instituto de F\'\i sica de Cantabria, Consejo Superior de
     Investigaciones Cient\'\i ficas-Universidad de Cantabria,
     Santander, Spain\\
   $^2$Departamento de F\'\i sica Moderna, Universidad de Cantabria,
     Avda. Los Castros s/n, 39005 Santander, Spain\\ 
   $^3$Department of Physics. Nuclear \& Astrophysics Laboratory,
        Keble Road, Oxford OX1 3RH, UK\\}

\maketitle\abstracts{
In this work we show that combining different cluster data sets is a 
powerful tool to constrain both, the cosmology and cluster properties. 
We assume a model with 9 parameters and fit them to 5 cluster data sets. 
From that fit, we conclude that only low density universes are 
compatible with the previous data and we also get some interesting 
conclusions on the rest of parameters.}

\section{Introduction}
Our main purpose when we started this work was to perform realistic 
simulations of the Sunyaev-Zel'dovich effect (SZE) in order to check 
different substraction techniques applied to this effect. 
By realistic we understand that our simulations must take into account 
first, the cluster population; that is, how is the cluster distribution in 
the mass-redshift space. Second, we must be able to describe cluster 
properties such as the scalings temperature-mass ($T-M$) and X-ray 
luminosity-mass ($L_x-M$). These descriptions of the cluster population and 
cluster scalings must be realistic in the sense that both descriptions must 
be in agreement with recent cluster data. \\
The idea of this work is to find such a realistic description of the clusters 
(population + intrinsic properties) by comparing our model with recent 
X-ray and optical data and fitting the free parameters of the model to 
the data in order to search for the best fitting values. \\

\noindent
Several authors have used different cluster data sets in an attempt to  
constrain the cosmology. The usuall procedure is, starting from the 
Press-Schechter (PS) mass function, fitting the experimental 
mass function or by using a given 
(that is fixed or non-free parameter) $T-M$ or $L_x-M$ relation construct 
some of the other cluster functions (temperature, X-ray luminosity or flux 
functions) and then compare with the corresponding data sets. 
This can be a dangerous process for two reasons. First, when considering 
just one data set one ensures that his best fitting model is compatible 
with that data set but it can be inconsistent with others. We have 
observed this behaviour in many cases. For instance some of the models 
inside the $68 \%$ confidence level region in the $\Omega-\sigma_8$ space 
in Bahcall \& Fan (1998) are inconsistent with, for example, the luminosity 
function of Ebeling et al. (1997). This point suggests that a consistent 
analysis should take into account the information coming from different 
experiments to avoid these incompatibilities. 
The second problem comes when some authors fix the $T-M$ 
 or $L_x-M$ relations. The scatter in this correlations is large enough to 
introduce important uncertainties in the final result 
(Voit \& Donahue (1998)). When these authors fix some of these relations, 
they probably are introducing a sistematic bias in their conclusions due 
to the fact that they suppose that the virial relation, for instance, is 
a good enough aproximation everywhere. 
We want to avoid all those uncertainties by fitting different data sets 
simultaneously without doing any assumption about the cosmology and 
the $T-M$ and  $L_x-M$ relations. 
\section{Data}
In this section we present the five different data sets we have used in 
our fit. The first data set is the cluster mass function given in 
Bahcall \& Cen. (1993). To take into account evolutionary effects of 
the cluster mass function we use the evolution of 
the mass function given by Bahcall \& Fan 1998. With the $T-M$ and $L_x-M$ 
relations we can build the cluster temperature, X-ray luminosity and 
flux functions. For the temperature function we use the one given by  
Henry \& Arnaud (1991). For the X-ray luminosity function, that  
of Ebeling et al. 1997 and finally for the X-ray cluster flux function, the 
one determined by Rosati et al. 1998 for low-flux clusters, and the 
one obtained by De Grandi et al. (1999) for high-flux clusters.
With all these data curves and with our model we are now able to fit the model 
and say something about its free parameters. The results will be 
robust and consistent in the sense that we did not make any assumption 
about the cosmology (we did not fix any cosmological parameter) 
nor about the intrinsic cluster properties ($T-M$ and $L_x-M$).
\section{The model}
Our model consists of two parts. First the description of the cluster 
population and second the intrinsic cluster properties. For the cluster 
population we assume the standard Press-Schechter (PS) formalism (Press \& 
Schechter 1974). 
This formalism depends on three parameters: the density of the universe 
$\Omega$, the amplitude of the power spectrum in units 
of $\sigma_8$ and finally the shape parameter of the power spectrum 
$\Gamma$. In this work we have only considered low density models with 
$\Lambda = 0$. In a subsequent paper we will also include in our analisis 
flat $\Lambda$CDM models.\\
The PS approach is supported by N-body numerical simulations which do show 
a good agreement with the PS parametrization (Lacey \& Cole 1994, 
White et al. 1993, Efstathiou et al. 1988, etc).\\ 
With the PS formalism we know the comoving number density of 
clusters with $M \in [M,M+dM]$ and at a given redshift. 
This will allow us to distribute the clusters in the $M-z$ space for 
a given solid angle and cosmology ($\Omega,\sigma_8,\Gamma$). 
What we need now is the second part of the model, the intrinsic cluster 
properties. Basically what we need is the $T-M$ relation for which the 
virial relation is usually assumed (Navarro et al. 1995, Bryan \& Norman 
1998). One problem is that it is not clear to what extend the virial relation 
is true for high or even intermediate redshift. In this work we have decided 
to consider this scaling as a free parameter relation:
\begin{equation}
  T_{gas} = T_0 M_{15}^{\alpha}(1 + z)^{\psi}
\end{equation}
\noindent
where $T_0, \alpha$ and $\psi$ are our three free parameters. 
$M_{15}$ is the cluster mass in $h^{-1} 10^{15} M_{\odot}$ units.
With the $T-M$ relation and the PS mass function we can build the 
temperature function, simply by doing:
\begin{equation}
 \frac{dN(T,z)}{dT} =\frac{dN(M,z)}{dM}\frac{dM}{dT}  
\end{equation}
Our aim is to use as many data sets as possible and so, we also want to 
include in our analysis other relevant data sets like the X-ray 
luminosity and flux functions. 
To do that we need another scaling law, the $L_x-M$ relation 
in order to relate the mass function with the X-ray luminosity and flux 
functions similarly as we did with the temperature function in eq. (2).
As in the $T-M$ relation, the $L_x-M$ is not well established yet and for this 
reson we have adopted a parametrization similar to that given in eq. (1):
\begin{equation}
  L_x^{Bol} = L_0 M_{15}^{\beta}(1 + z)^{\phi}
\end{equation}
\noindent
where we added three additional free parameters: $L_0, \beta$ and $\phi$. 
From this relation is possible to build the $S_x-M$ relation. Just taking, 
\begin{equation}
  S_x^{Bol} = \frac{L_x^{Bol}}{4 \pi D_l(z)^2 }
\end{equation}
\noindent
Summarising, the final number of free parameters are 9 :
$\Omega, \sigma_8, \Gamma, T_0, \alpha, \psi, L_0, \beta$ and $\phi$.\\
We now want to play with these parameters and look for the best fitting 
model to the different data sets.
\section{Best fit}
\begin{figure}
   \begin{center}
   \epsfxsize=9.cm 
   \begin{minipage}{\epsfxsize}
         \epsffile{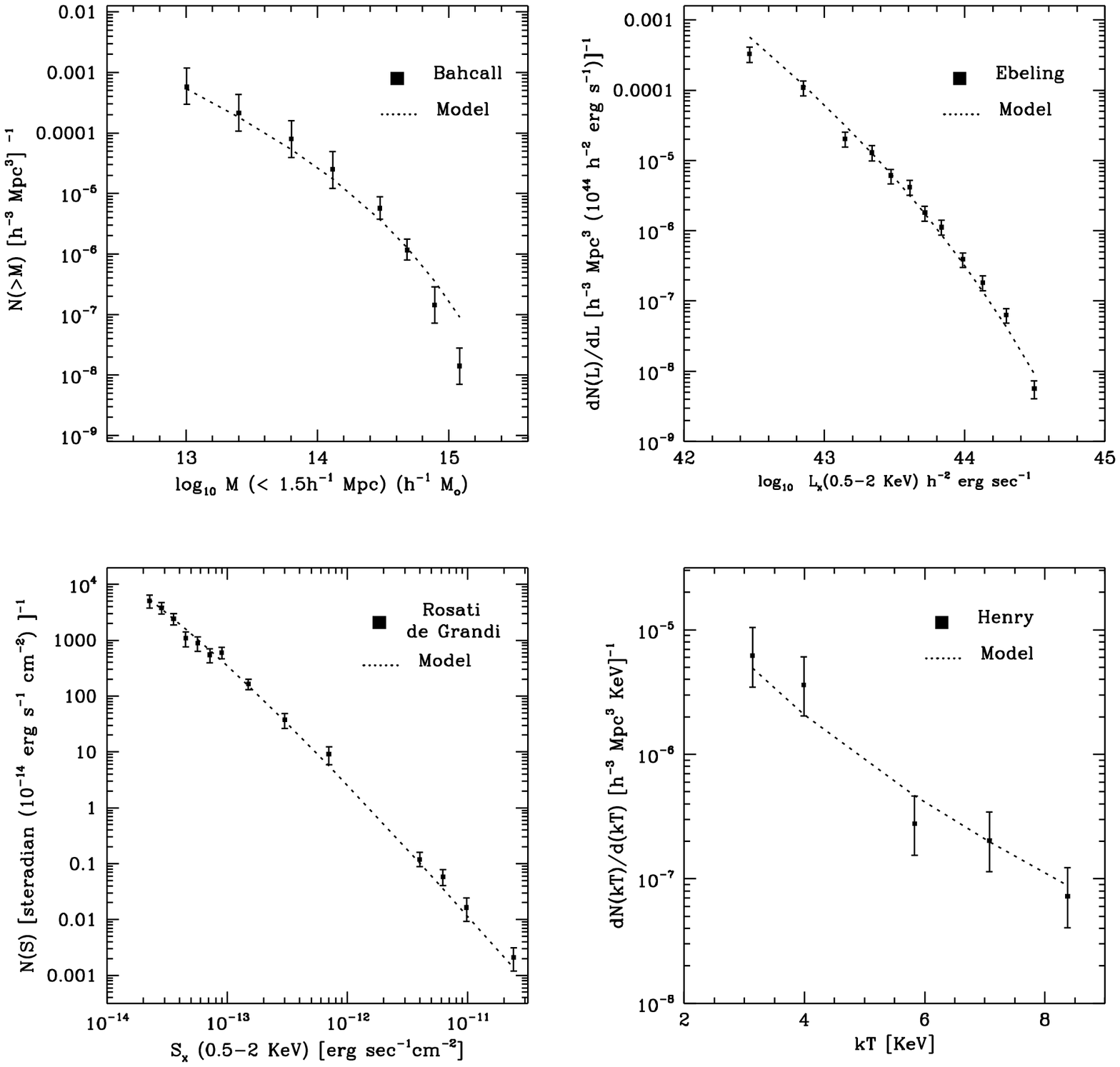}\end{minipage}
   \epsfxsize=5.cm 
   \begin{minipage}{\epsfxsize}
         \epsffile{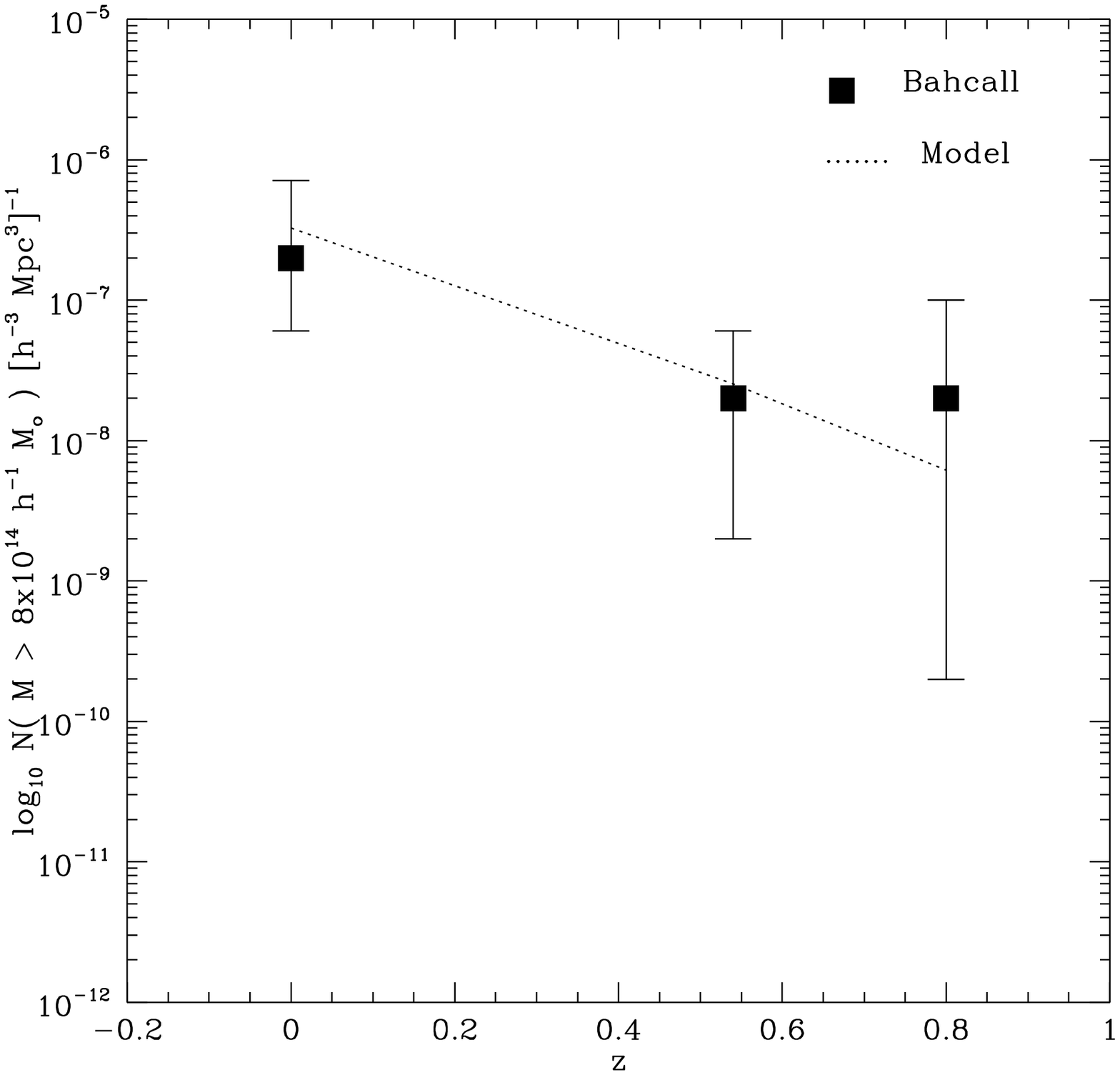}\end{minipage}
   \caption{Expected curves compared with the data for the best 
   CDM ($\lambda = 0$) model.}
   \end{center}
\end{figure}
\noindent
To fit the five data sets we must decide which estimator will pick up our 
best fitting model. 
Because of the scaling relations assumed between the mass and 
temperature ($T-M$) and the mass and luminosity in the X-ray band ($L_x-M$), 
then there must be some correlations among the five data sets predicted by 
the model. 
Therefore we should start by considering an estimator like the standard 
likelihood estimator.\\
In our case, the model depends on 9 free parameters and if we consider a 
grid of, let's say 5 values per parameter, then we should compute the 
correlation matrix for $5^9 \sim $ 1 million different models. This process 
would take many years. A faster technique would require a search method 
that avoids exploring all the parameter space. 
This can be the solution if we are interested just in the best model 
but we want also the error bars of our model, or in other words the 
marginalized probability distribution of the parameters. In order to do 
that, we need to know the probability in a given regular grid.\\
To simplify the problem, the most simple approach 
is to consider as our estimator the standard $\chi^2_{joint}$:
\begin{equation}
 \chi_{joint}^2 = \chi_M^2 + \chi_{M(z)}^2 + + \chi_{T}^2 +  
                  + \chi_{L_x}^2 + \chi_{S_x}^2 
\end{equation}
\noindent
where $\chi_x^2$ represents the corresponding ordinary $\chi^2$ for the 
five different data sets and we are supposing that the correlation matrix 
is in this case diagonal.\\
By doing this, we know that we are forgetting the correlations between 
the curves and that there will be some bias in our estimation. For this 
reason we want to check other more ellaborated estimators.\\
We have considered as a second estimator of the best model the next one 
based on bayesian theory (Lahav et al, 2000);
\begin{equation}
 -2 ln P_L = \chi_{L}^2  
\end{equation}
\noindent
where, 
\begin{equation}
 \chi_{L}^2 = \sum_i^5 N_i ln(\chi_i^2) 
\end{equation}
\begin{table}
\begin{center}
\caption{Best model.\label{fgastable}}
\begin{tabular}{lcccc}
   Parameter  & & \\
   \hline
   $\sigma_8$     		      & $ 0.8   $\\
   $\Gamma$      		      & $ 0.1   $\\
   $\Omega$      		      & $ 0.3   $\\
   $T_0 (10^8 K)$ 		      & $ 1.0   $\\
   $\alpha$       		      & $ 0.7   $\\
   $\psi$        		      & $ 1.0   $\\
   $L_0 (10^{45} h^{-2} erg/s)$       & $ 0.6   $\\
   $\beta$  	 		      & $ 3.1   $\\
   $\phi$			      & $ 2.0   $\\
\end{tabular}
\end{center}	
\end{table}
In this estimator $\chi^2$ is the same as before and $N_i$ represents the 
number of data points for the data set $i$.
The authors have shown that this estimator is apropiate for the case when 
different data sets are combined together, as is our case. The factor $N_i$ 
plays the role of a weight factor. Those data sets with more measures (more 
data points) are considered as more realistic.\\
We have checked both estimators by doing a bias test. In this test we have 
simulated the data for a known model with the corresponding error bars. The 
input model was selected according to the criterium that it would be as 
close as possible to the data (for instance the model which minimises 
$\chi^2_{joint}$).
In the simulations we have taken into account all the characteristics of 
the data, that is, sky coverage, limiting flux, maximun redshift etc.
Then we compare each one of these realizations with the mean value of the 
different models and for each realization we get the best model using both 
estimators. \\
We have concluded from this test that the second estimator works  
better than the standard $\chi_{joint}^2$. There is still some bias with 
the second estimator but the agreement between the input model and the 
recovered one is very good. \\
Using the second estimator we have computed the probability distribution 
in our 9 parameter space. We have used a grid with about 1 million different 
models and for each of them we have computed its $P_{L}$ (eq. 6). 
Knowing that, we can obtain the best model (maximum probability, see table 1).
\section{Conclusions}
In this work we have shown that combining different data sets and using the 
Lahav's et al. estimator is a powerful tool to constrain the cosmology 
and cluster parameters. In a future paper (Diego et al, in preparation) 
we will present the full analysis taking into account both, 
open ($\Lambda = 0$) and flat ($\Lambda > 0$) CDM models. 
The marginalized probability distributions for the parameters of the 
models will be also included.\\
Additional data coming from high redshift clusters (CHANDRA, 
XMM-Newton, PLANCK) will improve this result.\\
Particularly interesting is the work that can be done with future CMB surveys.
The PLANCK satellite will explore the whole sky at nine different 
frecuencies (from 30 Ghz to 800 Ghz) and with resolutions between 5 arcmin 
and 30 arcmin. At these frecuencies and with those resolutions we have shown 
(Diego et al. in preparation) that several clusters are expected to 
be observed at high redshift ($z > 2$) through the Sunyaev-Zel'dovich effect. 
The information that these clusters will provide will be decisive 
to definitely exclude many models. 

\section*{Acknowledgments}
We would like to thank to Piero Rosati for kindly providing his data for the 
differential flux function. JMD acknowledges the DGES for a fellowship.
JMD, EM, JLS, \& LC thanks CFPA/Astronomy Dept. Berkeley for the facilities 
given during this work.\\

\section*{References}

\end{document}